\UseRawInputEncoding

\documentclass[journal]{IEEEtran}
\usepackage{amsmath,amsfonts,bm,bbm}
\usepackage{amsthm}
\usepackage{algcompatible}
\usepackage{algorithm}
\usepackage{array}
\usepackage{subfigure}
\usepackage[caption=false,font=normalsize,labelfont=sf,textfont=sf]{subfig}
\usepackage{textcomp}
\usepackage{booktabs}
\usepackage{stfloats}
\usepackage{url}
\usepackage{verbatim}
\usepackage{multirow}
\usepackage{makecell}

\usepackage{graphicx}
\usepackage{soul, color,xcolor}

\usepackage{cite}
\usepackage{upgreek}

\begin{document}
\bstctlcite{IEEEexample:BSTcontrol}
    \title{F2PAD: A General Optimization Framework for Feature-Level to Pixel-Level Anomaly Detection} 
  \author{Chengyu Tao, Hao Xu, and Juan Du ~\IEEEmembership{Member,~IEEE}
  
  \thanks{C. Tao is with the Academy of Interdisciplinary Studies, The Hong Kong
University of Science and Technology, Hong Kong SAR, China (e-mail: ctaoaa@connect.ust.hk).}
    \thanks{H. Xu is with the Smart Manufacturing Thrust, The Hong Kong University of Science and Technology (Guangzhou), Guangzhou, China (e-mail: hxu726@connect.hkust-gz.edu.cn).}
\thanks{J. Du is with the Smart Manufacturing Thrust, The Hong Kong University of Science and Technology (Guangzhou), Guangzhou, China, the Department of Mechanical and Aerospace Engineering, the Academy of Interdisciplinary Studies, The Hong Kong University of Science and Technology, Hong Kong SAR, China, and also the Guangzhou HKUST Fok Ying Tung Research Institute, Guangzhou, China (e-mail: juandu@ust.hk).}
  
  }

\maketitle

\begin{abstract}

Image-based inspection systems have been widely deployed in manufacturing production lines. Due to the scarcity of defective samples, unsupervised anomaly detection that only leverages normal samples during training to detect various defects is popular. Existing feature-based methods, utilizing deep features from pretrained neural networks, show their impressive performance in anomaly localization 
and the low demand for the sample size for training.
However, the detected anomalous regions of these methods always exhibit inaccurate boundaries, which impedes the downstream tasks. This deficiency is caused: (\romannumeral 1) The decreased resolution of high-level features compared with the original image, \textit{and} (\romannumeral 2) The mixture of adjacent normal and anomalous pixels during feature extraction. To address them, we propose a novel unified optimization framework (F2PAD) that leverages the \underline{F}eature-level information \underline{to} guide the optimization process for \underline{P}ixel-level \underline{A}nomaly \underline{D}etection in the inference stage. The proposed universal framework can enhance various feature-based methods with limited assumptions. Case studies are provided to demonstrate the effectiveness of our strategy, particularly when applied to three popular backbone methods: PaDiM, CFLOW-AD, and PatchCore. \textit{Our source code and data will be available upon acceptance.}

\end{abstract}


%
\IEEEpeerreviewmaketitle


\section{Introduction}
\label{sec:introcution}

\IEEEPARstart{A}utomated visual inspection of products by deploying cost-efficient image sensors has become a critical concern in industrial applications in recent years, ranging from microscopic fractions and crack detection on electrical commutators \cite{Tabernik2019JIM} to fabric defect detection on textiles \cite{silvestre2019public}. In realistic production lines, defective products are always scarce. Moreover, all possible types of defects that occurred during production are impossible to know beforehand \cite{Bergmann_2019_CVPR}. In this scenario, only utilizing normal (non-defective) samples to train modern machine learning models for defect detection becomes essential. Therefore, unsupervised anomaly detection has become a mainstream setting in practical applications.

\begin{figure}[!t]
\centering 
\includegraphics[width=3.2in]{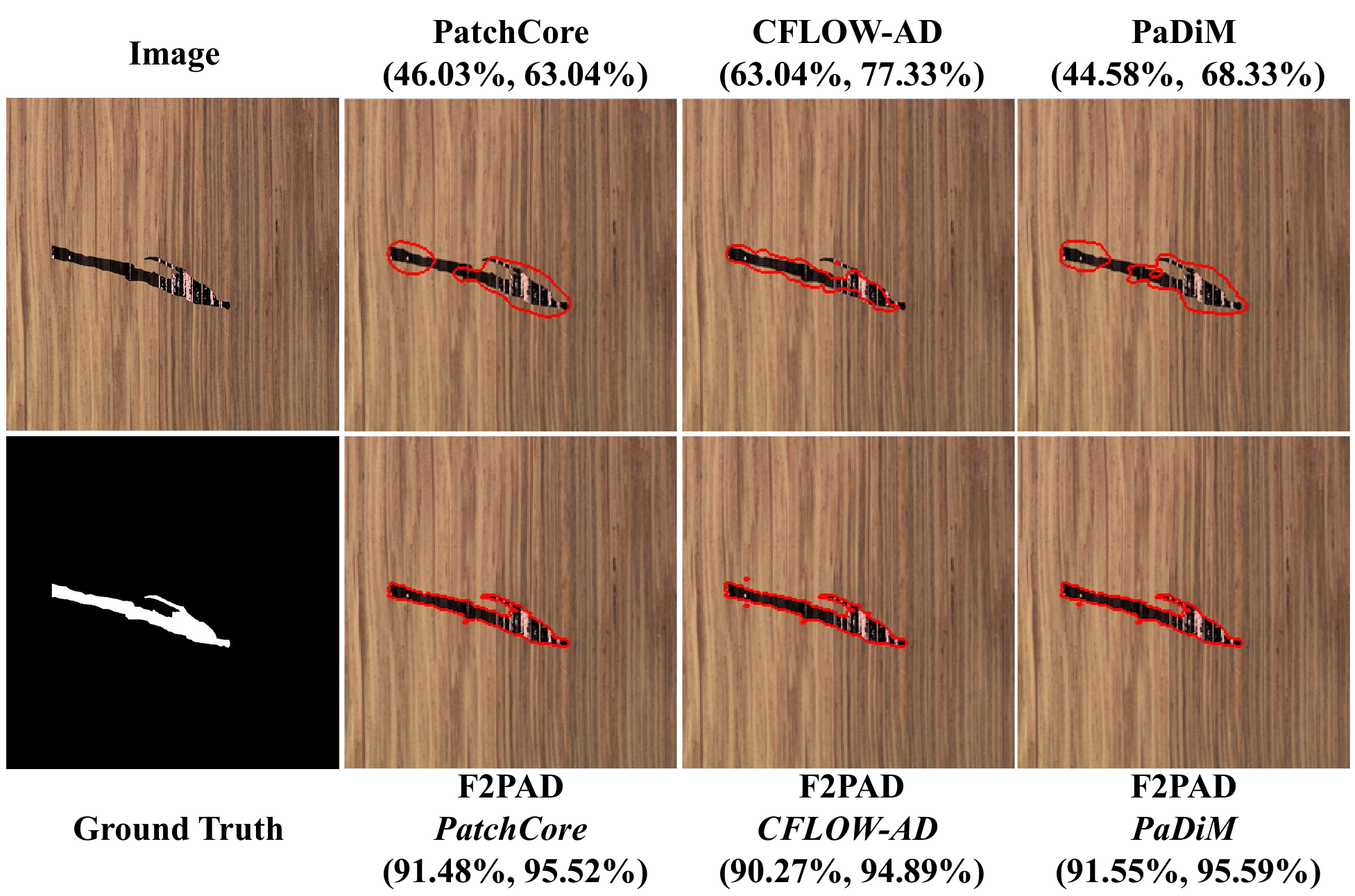}
\caption{An example of F2PAD's anomaly detection results. \textbf{Top row}: Test image and the anomaly detection results (Intersection over Union (IOU), DICE coefficient) of three popular feature-based methods, i.e., PatchCore \cite{roth2022towards}, CFLOW-AD \cite{gudovskiy2022cflow}, and PaDiM \cite{defard2021padim}. \textbf{Bottom row}: Ground truth and the results of F2PAD when applied to the above three methods. Our F2PAD significantly enhances the original PatchCore, PaDiM, and CFLOW-AD methods with higher anomaly segmentation accuracy, demonstrated by the improved IOU and DICE metrics.} 
\label{fig: intro}
\end{figure}

Besides determining whether a sample is anomalous or not, it is also essential to precisely localize and segment the defective regions due to the following reasons: (\romannumeral1) It enables the characterization of surface defects, such as shape, size, and orientation, which contributes to the evaluation of product quality. (\romannumeral2) It benefits the subsequent defect classification and recognition, as the models could focus more on the discriminative regions. An example is provided in Fig. \ref{fig: intro}, where the target defective region is masked as white in the ground truth (GT).

Current unsupervised image anomaly detection methods can be mainly categorized into two groups \cite{liu2023deep}: (1) \textit{Image reconstruction-based methods}. These methods deploy various neural network architectures for the reconstruction of an input image, such as Autoencoders (AE) \cite{bergmann2018improving,zavrtanik2021draem} and Generative Adversarial Networks (GAN) \cite{song2021anoseg}. Therefore, the per-pixel comparison between the input and reconstructed images can localize anomaly. (2) \textit{Feature-based methods}. These methods mainly rely on deep features extracted from neural networks pretrained on large-scale image databases like ImageNet\cite{deng2009imagenet}. The collected features of normal samples can be modeled in different ways. Then, associated distance metrics are defined to describe the normality of test features \cite{defard2021padim,rippel2021modeling,rudolph2021same,gudovskiy2022cflow}. Section \ref{sec: review} provides a more detailed literature review.

To guarantee the high-fidelity reconstruction of normal regions at the pixel level, {image reconstruction\allowbreak-based} methods require sufficient normal samples. In contrast, owing to the great expressiveness of deep features even for unseen manufacturing parts in industries, 
{feature\allowbreak-based} methods have less demand for the sample size of the training dataset. However, though the existence and location of anomalies can be identified well by the feature-based methods, the boundaries of detected anomalies are always inaccurate \cite{bergmann2020uninformed,tao2022deep,mou2023paedid}. To illustrate this point, the first row of Fig. \ref{fig: intro} visualizes the unsatisfactory anomaly segmentation results of PatchCore \cite{roth2022towards}, CFLOW-AD \cite{gudovskiy2022cflow}, and PaDiM \cite{defard2021padim} with relatively low evaluation metrics.

To explain the above common deficiency in feature-based methods, we summarize the two issues as follows:  
\begin{itemize}
 \item \textbf{Issue 1}: \textit {Decreased resolution of the feature map}. The popular choices for pretrained networks are Convolutional Neural Networks (CNN). With the multi-scale pyramid pooling layers, the feature map constructed from a deeper layer has a smaller resolution than the previous layers. For instance, in \cite {defard2021padim}, the resolution of an input image is $224 \times224$ (height $\times$ width), while the sizes of feature maps extracted from the $1\textrm{st}$ and $3\textrm{th}$ layers of the ResNet18 \cite {he2016deep} are $56\times56$ and $14\times14$, respectively. In addition, the patch representation, which is spatially downsampled from the original image, is adopted in Vision Transformers (ViTs) \cite{dosovitskiy2020image} and the self-supervised feature learning \cite{yi2020patch}. Since anomaly detection is conducted at the feature (patch) level, per-pixel accuracy cannot be achieved.
 
 \item \textbf{Issue 2}: \textit {Mixture of adjacent normal and anomalous pixels}. To extract a specific feature, all adjacent pixels falling within its receptive field (or the associated patch) are considered. Consequently, features near anomaly boundaries aggregate information from both normal and anomalous pixels. These features are often confusing and difficult to discriminate accurately, resulting in unclear anomaly segmentation.
\end{itemize}

This paper aims to tackle the above two issues concurrently, enabling accurate anomaly segmentation in feature-based methods. To achieve this, we propose a novel optimization framework for Feature-level to Pixel-level Anomaly Detection (F2PAD). Specifically, we can decompose an arbitrary input image into two \textit{unknown} parts, i.e., a \textit{non-defective image} plus an \textit{anomalous part}, which have the same resolutions as the original image. Then, an optimization model is proposed, which encourages the non-defective image to produce normal features. In this way, the anomaly is directly indicated by the \textit{estimated} anomalous part, which \textit{remains the original resolution}. Furthermore, feature extraction is performed solely on the \textit{estimated} non-defective image \textit{without the mixture of different types of pixels}. The above properties directly avoid the aforementioned issues. To end up, our main contributions are summarized as follows:

 \begin{itemize}
 \item We propose a general optimization framework to enhance a variety of feature-based methods for precise pixel-level anomaly segmentation in industrial images, with examples of PatchCore \cite{roth2022towards}, CFLOW-AD \cite{gudovskiy2022cflow}, and PaDiM \cite{defard2021padim}.

\item We propose a novel penalized regression model that concurrently estimates the non-defective image and anomalous part. The loss function originates from an established feature-based approach, and we incorporate two regularization terms: a sparsity-inducing penalty for identifying anomalies and a pixel prior term to enhance non-defective image recovery.

\item We design an optimization algorithm for the above regression problem with a novel local gradient-sharing mechanism, which empirically shows better optimization results.

\end{itemize}

The remainder of this paper is organized as follows. Section \ref{sec: review} reviews the literature about industrial image anomaly detection. Section \ref{sec: review} discusses the formulation of our proposed framework. Then, case studies including various categories of realistic products are provided in Section \ref{sec: experiment} to validate our methodology. Finally, Section \ref{sec: conclusion} summarizes this paper.

\section{Related Work}
\label{sec: review}

This section reviews the following two categories of industrial image anomaly detection methods in the literature.

\subsection{Image Reconstruction-Based Methods}
\label{sec: review reconstruction}
Reconstruction-based image anomaly detection methods aim to reconstruct the normal samples accurately during training, assuming imperfect reconstruction of defective regions for an anomalous image during inference. The pixel differences between the input and reconstructed images can be directly transformed into anomaly scores and further thresholded as a binary anomaly mask. Therefore, the key research goal is to pursue a more accurate reconstruction. Autoencoder (AE) is the most prevalent architecture for image reconstruction \cite{wang2020image, xu2023ano}. In AEs, a common choice of the backbone network is CNN. Vision transformers were introduced in  \cite{lee2022anovit,pirnay2022inpainting} to capture more global structural information. Besides the AEs, GANs \cite{song2021anoseg,yang2021memory} and diffusion models \cite {wyatt2022anoddpm,teng2022unsupervised} are also popular choices. 

Instead of focusing on the architecture of the reconstruction network, there are various strategies to improve the reconstruction accuracy. For example, the structure similarity index measure (SSIM) was introduced in \cite{bergmann2018improving} to incorporate the perceptual difference of anomaly. 
Additionally, to address the deterioration of image reconstruction caused by anomalies, \cite{Dehaene2020Iterative,mou2023rgi} enforced the reconstructed image on the learned manifold. Alternatively, \cite{zhou2022memorizing} considered the correspondence between structure and texture.

Nevertheless, sufficient normal samples are necessary for pixel-wise reconstruction in the image reconstruction-based methods \cite{mou2023paedid}.

\subsection{Feature-Based Methods}
\label{sec: review feature}

Utilizing features with higher-level information avoids the necessity of per-pixel reconstruction. Generally, feature-based methods consist of two subsequent steps \cite {tao2022deep}: (\romannumeral1) Feature extraction. It can be achieved by employing various CNNs and ViTs pretrained on large-size image databases or can be learned in a self-supervised way with patch representation \cite{yi2020patch}. (\romannumeral2) The modeling of normal features. Most methods in the literature focus on this step, which will be comprehensively reviewed in this subsection. After this step, anomaly scores are assigned to individual features and then simply interpolated to yield pixel-level results.

PaDiM \cite{defard2021padim} adopted the multivariate Gaussian distributions to model the normal features. To allow more complicated probabilistic distributions, DifferNet \cite{rudolph2021same} adopted a neural network, i.e., normalizing flow, to estimate the density of features. Furthermore, CFLOW-AD \cite{gudovskiy2022cflow} improved the DifferNet by modeling the conditional probabilities of features on their locations. Similarly, a 2D normalizing flow was proposed in \cite{yu2021fastflow} to retain the spatial relationship between features, while the multi-scale features were considered in \cite{zhou2024msflow}.
The memory bank-based method that collects all normal features was first proposed in \cite{cohen2020sub}, while the average distance of a test feature to its k-nearest neighbors was regarded as the anomaly score. To improve the inference speed, the coreset technique was adopted in PatchCore\cite{roth2022towards} by greedily selecting the farthest features to construct a reduced memory bank.
Similarly, \cite{xu2022discriminative} proposed a gradient preference strategy to determine the most significant features for anomaly detection.
In addition,  \cite{Massoli2022Mocca} encouraged the normal features within compact spheres. 
 Teacher-student networks were also popular in literature \cite{ salehi2021multiresolution}, which effectively transfer normal features from the teacher network to the student model but fail for anomalous features.

One issue of the above methods is the decreased resolution of the feature map, as stated in \textbf{Issue 1}. To tackle it, the work \cite{bergmann2020uninformed} learned dense features in the image size by using the multipooling and unwarping strategy in \cite{bailer2018fast}. However, it is only applicable to local patch descriptions by CNNs. Furthermore, the anomaly scores were defined as the gradients of loss function w.r.t the input image in \cite{salehi2021multiresolution}. Therefore, feature-level scores are no longer required. However, this strategy may not achieve satisfactory results for general methods in practice \cite{Dehaene2020Iterative}. Finally, the works \cite{bae2023pni} and \cite{Xing2024adps} adopted additional segmentation networks to refine the initial anomaly map. However, these methods require more samples to train the segmentation networks. In addition, the mixture of normal and anomalous pixels described in \textbf{Issue 2} remains unsolved in the above methods \cite{bergmann2020uninformed,salehi2021multiresolution,bae2023pni}. 

In summary, current literature lacks a general framework to deal with \textbf{Issues 1\&2} discussed in Section \ref{sec:introcution} simultaneously in feature-based methods. Our proposed F2PAD can achieve the above goal with limited assumptions, which will be discussed in detail in the following Section \ref{sec: methodology}. 

\section{Methodology}
\label{sec: methodology}

We first clarify our objective and formulate the problem as a penalized regression problem by our F2PAD in Section \ref{sec: overview}. Then, Sections \ref{sec: loss} and \ref{sec: regularization} show the details about the loss functions and regularization terms respectively. In addition, we propose a novel optimization algorithm in Section \ref{sec: algorithm}. Finally, the implementation details are present in Section \ref{sec: implementation}.

\subsection{Overview}
\label{sec: overview}

\subsubsection{Preliminary}
Given an input RGB image $\mathbf{x} \in \mathbb{R}^{h\times w\times3}$ with height $h$ and width $w$, our goal is to obtain a binary anomaly mask $\mathbf{m} \in \{0,1\}^{h\times w}$, where $\mathbf{m}_{i,j} = 1$ indicates an anomalous pixel in row $i$ and column $j$, and normal otherwise. 

For the feature-based methods, a feature extractor $\mathbf{\phi}_k$ is utilized to extract the feature map  $\mathbf{f}_k = \mathbf{\phi}_k(\mathbf{x}) \in \mathbb{R}^{h_k\times w_k \times c_k}$ in layer $k$, where $h_k$, $w_k$, and $c_k$ are height, width, and the number of feature channels, respectively. Then, anomaly score $s_{k,i,j} \in \mathbb{R}$ assigned to the feature $\mathbf{f}_{k,i,j} \in \mathbb{R}^{c_k}$ is given by the scoring function $s_{k,i,j} = \mathbf{\psi}_k(\mathbf{f}_{k,i,j})$. Notably, \textbf{Issue 1} in Section \ref{sec:introcution} indicates that the resolution $(h_k,w_k)$ is always smaller than $(h,w)$. $\mathbf{\phi}_k$ is learned during training and fixed for inference.  We use $\nabla$ to denote the gradient.

\subsubsection{Assumptions}
\label{sec: assumptions}
To allow the formulation of our F2PAD, we make the following two assumptions: 
\begin{itemize}
    \item The (approximated) gradients of feature extractor $\mathbf{\phi}_k$ and scoring function $\mathbf{\psi}_k$ exist. For example, $\mathbf{\phi}_k$ defined by neural networks in almost all existing works, e.g., CNNs and ViTs, meet the above requirement that enables backpropagation (BP). Furthermore, $\mathbf{\psi}_k$ is always defined as probabilistic metrics, e.g., the log-likelihood, or distance metrics, such as the $l_2$ and Mahalanobis distances. Modern deep learning libraries can automatically compute the gradients (approximately) of common mathematical operations, such as the $\mathrm{autograd}$ module in PyTorch \cite{paszke2017automatic}. Therefore, a variety of methods support this assumption.

    \item The anomalies are locally distributed. The common surface defects, including holes, dents, cracks, and scratches, show the above property. In addition, this assumption is also adopted in various anomaly detection works \cite{tao2023pointsgrade,mou2023rgi,cao20243d,Dehaene2020Iterative}. 
 
\end{itemize}

\subsubsection{Overall Framework}
Motivated by the decomposition ideas \cite {Dehaene2020Iterative,yan2017anomaly}, we assume an additive model as follows:
\begin{equation}
 \mathbf{x} = \mathbf{n} + \mathbf{a},
\label{eq: decomposition}
\end{equation}
where $\mathbf{n}, \mathbf{a}\in \mathbb{R}^{h\times w\times3}$ are the non-defective image and the anomalous part respectively. The relationship between the mask $\mathbf{m}$ and  $\mathbf{a}$ is that 
\begin{equation}
    \mathbf{m}_{i,j}=1 \ \mathrm{iff} \ \mathbf{a}_{i,j} \neq0.
\label{eq: m}
\end{equation}

Then, the anomaly segmentation problem in this paper is to estimate $\mathbf{n}$ and $\mathbf{a}$ in a unified model. Specifically, the separation of $\mathbf{n}$ and $\mathbf{a}$ in Eq. \eqref{eq: decomposition} indicates that there is no anomaly in the non-defective image $\mathbf{n}$. Then, we can define a loss function $l_n(\mathbf{n})$ to encourage small anomaly scores $\{s_{k,i,j}\}$ for the features extracted from $\mathbf{n}$. Concurrently, a sparsity-inducing penalty $l_a(\mathbf{a})$ is introduced to promote the locality of anomalies, as assumed in Section \ref{sec: assumptions}. Since $l_n(\mathbf{n})$ is related to high-level features, another regularization term, i.e., the pixel prior $l_p(\mathbf{n})$, is necessary to enhance the recovery of $\mathbf{n}$. Finally, the penalized regression problem with objective function $F(\mathbf{n},\mathbf{a})$ is formulated as follows:
\begin{equation}
 \mathop{\min}\limits_{\mathbf{n},\mathbf{a},\mathbf{x}= \mathbf{n} + \mathbf{a}} F(\mathbf{n},\mathbf{a})=l_n(\mathbf{n}) + \alpha l_p(\mathbf{n}) + \beta l_a(\mathbf{a}).
\label{eq: model}
\end{equation}

\subsubsection{Discussion} There are the following properties in the problem \eqref{eq: model} that 
contribute to the goal of this paper: 
\begin{itemize}
    \item For \textbf{Issue 1}, $\mathbf{a}$ remains the same resolution as $\mathbf{x}$, indicating that the pixel-level detection can be directly given by solving the problem \eqref{eq: model}. 

    \item For \textbf{Issue 2}, feature extraction is solely performed on $\mathbf{n}$, as shown in loss $l_n(\mathbf{n})$, which is not influenced by the appeared anomalies. 

    \item The first assumption in Section \ref{sec: assumptions} means that the problem \eqref{eq: model} is practically solvable through gradient-based algorithms. 
\end{itemize}

The following subsections will show more details about the problem \eqref{eq: model}.

\subsection{Loss Functions}
\label{sec: loss}

This subsection shows three examples of loss functions $l_n(\mathbf{n})$ when applied to PatchCore \cite{roth2022towards}, CFLOW-AD \cite{gudovskiy2022cflow}, and PaDiM \cite{defard2021padim}, respectively.

\subsubsection{PatchCore}: Following \cite{roth2022towards}, we concatenate all feature maps in different layers into a single one $\mathbf{f} \in \mathbb{R}^{h_1\times w_1\times c}$, with $c = \sum_k c_k$, remaining the same size as the largest feature map $\mathbf{f}_1$. Given the memory bank $\mathcal{M}$ with $|\mathcal{M}|$ normal features collected from the training dataset,  the anomaly score $s_{i,j}$ of $\mathbf{f}_{i,j}$ is defined as the minimum distance to the memory bank:
\begin{equation}
s_{i,j} = \min_{\mathbf{f}_{m} \in \mathcal{M}} \|\mathbf{f}_{i,j}-\mathbf{f}_{m} \|^2_2. 
\label{eq: patchcore}
\end{equation}

We need to consider the whole $\mathcal{M}$ in Eq. \eqref{eq: patchcore}, which heavily influences the computational speed. In practice, we can only consider a small candidate set $\mathcal{M}_{i,j} \in \mathcal{M}, |\mathcal{M}_{i,j}| \ll |\mathcal{M}|$ for each $\mathbf{f}_{i,j}$ to approximate $s_{i,j}$ as 
\begin{equation}
s^{\prime}_{i,j} = \min_{\mathbf{f}_{m} \in \mathcal{M}_{i,j}} \|\mathbf{f}_{i,j}-\mathbf{f}_{m} \|^2_2. 
\label{eq: approx patchcore}
\end{equation}

Based on Eq. \eqref{eq: approx patchcore}, we can define the final loss as the summation of scores over all pixels, i.e.,  $l_n(\mathbf{n}) = \sum_{i,j}s^{\prime}_{i,j}$. 

To construct $\mathcal{M}_{i,j}$, we assume that we have an initial estimate of $\mathbf{n}$ in \eqref{eq: model}, denoted as $\mathbf{n}^{(0)}$, which subsequently provides an initial $\mathbf{f}_{i,j}^{(0)}$. Then, we define $\mathcal{M}_{i,j}$ as the $|\mathcal{M}_{i,j}|$ nearest features of $\mathbf{f}_{i,j}^{(0)}$ in $\mathcal{M}$. Therefore, if $\mathbf{f}_{i,j}^{(0)}$ is not far from the optimal value, Eq. \eqref{eq: approx patchcore} can approximate well Eq. \eqref{eq: patchcore}. We will discuss how to access $\mathbf{n}^{(0)}$ in Section \ref{sec: initialization}. 

\subsubsection{CFLOW-AD}: In this method, to incorporate the positional information, the feature map $\mathbf{f}_k$ in layer $k$ is concatenated with $\mathbf{c}_k \in R^{h_k \times w_k \times 128}$, i.e., $\mathbf{f}_k = \mathbf{f}_k \oplus \mathbf{c}_k$. $\mathbf{c}_k$ is obtained by the position encoding with $\mathrm{sin}$ and $\mathrm{cos}$ harmonics \cite{gudovskiy2022cflow}. Notably, $\mathbf{c}_k$ is irrelevant to $\mathbf{n}$.

The log-likelihood $\log p(\mathbf{f}_{k,i,j}|\mathbf{\theta}_k)$ is modeled by a normalizing flow with parameters $\mathbf{\theta}_k$.
To speed up the optimization, we apply the $\mathrm{LogSigmoid}$ operation to the log-likelihood to define the anomaly score as:
\begin{equation}
 s_{k,i,j} = \log (\dfrac{p(\mathbf{f}_{k,i,j})}{(1+p(\mathbf{f}_{k,i,j})}). 
\label{eq: CFLOW-AD}
\end{equation}
Similar to PatchCore, we define $l_n(\mathbf{n}) = \sum_{k,i,j}s_{k, i,j}$.

\subsubsection{PaDiM}: PaDiM follows the same procedure of feature concatenation as PatchCore.  $\mathbf{f}_{i,j}$ follows a multivariate Gaussian distribution with mean $\mathbf{\mu}_{i,j} \in \mathbb{R}^{c}$ and covariance $\mathbf{\Lambda}_{i,j} \in \mathbb{R}^{c \times c}$. Then, the anomaly score $s_{i,j}$ is defined by the square of Mahalanobis distance \cite{defard2021padim} as follows:
\begin{equation}
 s_{i,j} = (\mathbf{f}_{i,j}-\boldsymbol{\mu}_{i,j})^T\mathbf{\Lambda}_{i,j}^{-1}(\mathbf{f}_{i,j}-\boldsymbol{\mu}_{i,j}).
\label{eq: PaDiM}
\end{equation}
We have $l_n(\mathbf{n}) = \sum_{i,j}s_{i,j}$.

\subsubsection{Discussion}
Since the above methods adopt neural networks as feature extractor $\mathbf{\phi}_k$, we  can easily access the gradients of $\mathbf{f}_{i,j}$ and $\mathbf{f}_{k,i,j}$ w.r.t. $\mathbf{n}$, i.e., $\nabla_{\mathbf{n}}\mathbf{f}_{i,j}$ and $\nabla_{\mathbf{n}}\mathbf{f}_{k,i,j}$ respectively. Moreover, Eqs. \eqref{eq: approx patchcore}, \eqref{eq: CFLOW-AD}, and \eqref{eq: PaDiM} show that it is able to derive the gradients $\nabla_{\mathbf{f}_{i,j}}s_{i,j}$ and $\nabla_{\mathbf{f}_{k,i,j}}s_{k,i,j}$.  Therefore, the first assumption in Section \ref{sec: assumptions} is valid for the above three methods.
Following the chain rule, we can acquire $\nabla_{\mathbf{n}}l_n(\mathbf{n})$. 

\subsection{Regularization Terms}
\label{sec: regularization}

\begin{figure*}[!ht] 
\centering 
\includegraphics[width=0.95\textwidth]{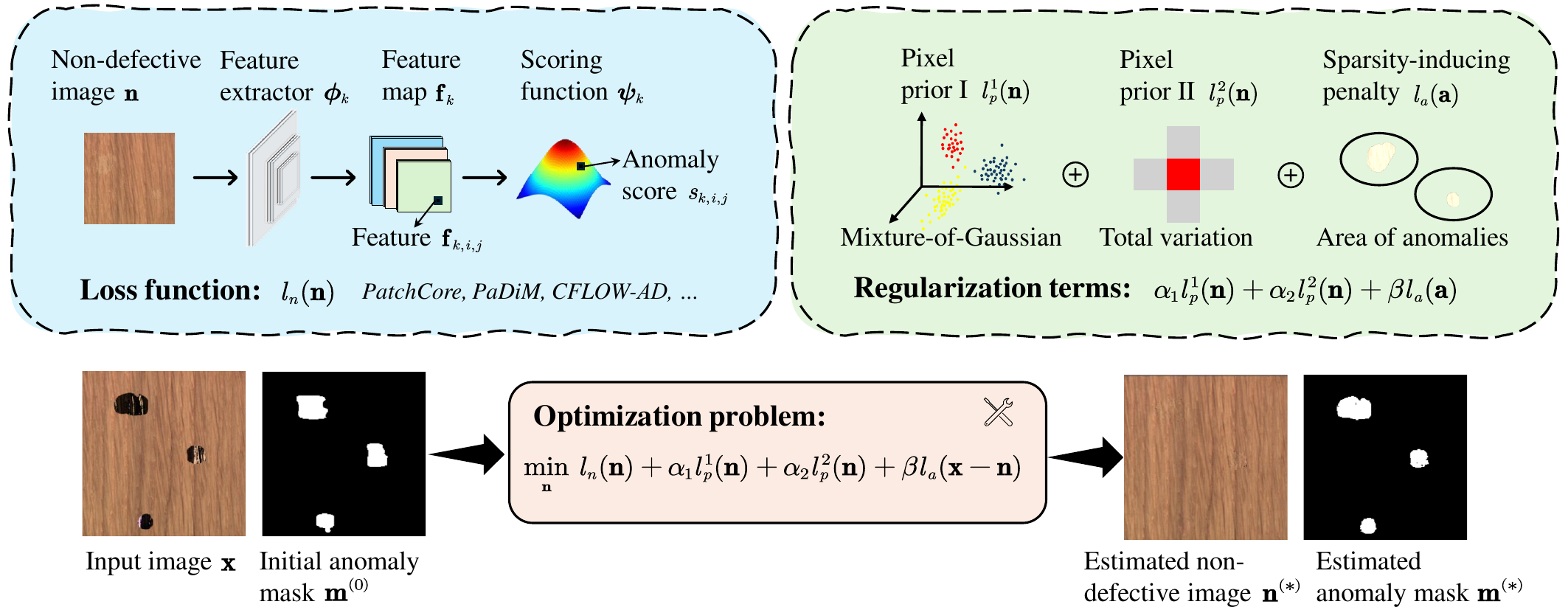} 
\caption{Schematic diagram of our F2PAD. We propose a novel optimization problem, which consists of the loss function $l_n(\mathbf{n})$ derived from feature-based methods, and three regularization terms $l_p^1(\mathbf{n})$, $l_p^2(\mathbf{n})$, and $l_a(\mathbf{a})$ as different roles. 
Given the input image $\mathbf{x}$ and initial anomaly mask $\mathbf{m}^{(0)}$ obtained by the original backbone methods (see Section \ref{sec: initialization}), we can accurately estimate the non-defective image $\mathbf{n}^{(*)}$ with mask $\mathbf{m}^{(*)}$ by solving the proposed optimization problem.} 
\label{fig: schematic diagram}
\end{figure*}

This subsection illustrates the motivation and specific definitions of $l_a(\mathbf{a})$ and $l_p(\mathbf{n})$.

\subsubsection{$l_a(\mathbf{a})$}
This term is used to promote local anomalies. From the definition of $\mathbf{a}$ in Eq. \eqref{eq: decomposition}, the number of anomalous pixels is given by the $l_0$ norm $\sum_{i,j}\|\mathbf{a}_{i,j}\|_0$. However, the $l_0$-based problem is NP-hard and the gradient does not exist, making it hard to efficiently solve the problem \eqref{eq: model}. Although the $l_1$ norm is a common practice \cite{mou2023rgi,Dehaene2020Iterative} as the convex relaxation of $l_0$ norm, it suffers the issue of inaccurate estimation in anomaly detection \cite{tao2023pointsgrade}. To address it, this paper adopts the nonconvex LOG penalty \cite{ke2021iteratively,tao2023pointsgrade} for $l_a(\boldsymbol{a})$ as follows:
\begin{equation}
 l_a(\mathbf{a}) = \sum_{i,j} \log(\sqrt{\|\mathbf{a}_{i,j}\|^2_2+ \epsilon} + \|\mathbf{a}_{i,j}\|_2),
\label{eq: sparse}
\end{equation}
where $\epsilon$ is a hyperparameter.

\subsubsection{$l_p(\mathbf{n})$}
\label{sec: pixel prior}
It is also important to accurately recover $\mathbf{n}$ in \eqref{eq: model}. However, the loss $l_n(\mathbf{n})$ is directly defined by the high-level information encoded by features, as discussed in Section \ref{sec: loss}. This indicates that the low-level pixel values cannot be uniquely determined. Consequently, the estimation of $\mathbf{a}$ will be influenced, leading to inaccurate detection results. To tackle it, we introduce the following two types of pixel priors.

Since the training dataset is available, the pixel value of a test image is expected to fall within some known ranges. Formally, we can model the prior distribution of a pixel. An image may contain the background and the foreground object, while the object comprises several components associated with different colors. Therefore, we assume that the pixel $\mathbf{n}_{i,j}$ follows a mixture-of-Gaussian (MOG) distribution:
\begin{equation}
\begin{split}
     p(\mathbf{n}_{i,j}) &= \sum_q r_q \cdot (2\pi)^{-3/2}|\mathbf{\Lambda}_q|^{-1/2} \\
      \cdot& \exp(-\dfrac{(\mathbf{n}_{i,j}-\boldsymbol{\mu}_q)^T\mathbf{\Lambda}_q^{-1}(\mathbf{n}_{i,j}-\boldsymbol{\mu}_q)}{2}),
\label{eq: mog}
\end{split}
\end{equation}
where $r_q \in \mathbb{R}$,  $\boldsymbol{\mu}_q \in \mathbb{R}^3$, and $\mathbf{\Lambda}_q \in \mathbb{R}^{3\times3}$ is the prior, mean, and covariance of the $q\mathrm{th}$ component of MOG, respectively. We adopt the algorithm \cite{attias1999variational} to estimate  $r_q$,  $\boldsymbol{\mu}_q$, and $\mathbf{\Lambda}_q$ by randomly sampling sufficient pixels from the training dataset. 

To construct  $l_{p}(\mathbf{n})$, we do not assign the priors $\{r_q\}$. Instead, we assume the equal probability of each component of MOG. This reduces the bias of converging to the most frequently occurring pixels in the training set. Combining Eq. \eqref{eq: mog} and similar to the Mahalabobis distance in Eq. \eqref{eq: PaDiM}, we can define a specification of $l_{p}(\mathbf{n})$ as:
\begin{equation}
    l_{p}^1(\mathbf{n}) = \sum_{i,j} \min_q (\mathbf{n}_{i,j}-\boldsymbol{\mu}_q)^T\mathbf{\Lambda}_q^{-1}(\mathbf{n}_{i,j}-\boldsymbol{\mu}_q),
    \label{eq: prior gmm}
\end{equation}
which focuses on the most likely component.

Furthermore, Eq. \eqref{eq: prior gmm} is defined by individual pixels, ignoring the spatial relationship between neighboring pixels. Since the images of industrial objects always exhibit piece-wise smoothness, we consider the total variation (TV) regularization as another term  $l_{p}^2(\mathbf{n})$ to promote the piece-wise smoothness  \cite{miyata2015inter}:
\begin{equation}
 l_{p}^2(\mathbf{n}) = \sum_{i,j}\|\mathbf{n}_{i+1,j}-\mathbf{n}_{i,j}\|_2 + \|\mathbf{n}_{i,j+1}-\mathbf{n}_{i,j}\|_2.
\label{eq: prior tv}
\end{equation}

With Eqs. \eqref{eq: prior gmm} and \eqref{eq: prior tv}, we can substitute $\mathbf{a} = \mathbf{x} - \mathbf{n}$ 
and rewrite the problem \eqref{eq: model} as:
\begin{equation}
    \min_{\mathbf{n}} F(\mathbf{n}) = l_n(\mathbf{n}) + \alpha_1 l_p^1(\mathbf{n}) + \alpha_2 l_{p}^2(\mathbf{n})  + \beta l_a( \mathbf{x} - \mathbf{n}),
\label{eq: final model}
\end{equation}
where  $\alpha_1$ and  $\alpha_2$ are the corresponding tuning parameters of $l_p^1(\mathbf{n})$ defined in Eq. \eqref{eq: prior gmm}$l_p^2(\mathbf{n})$ defined in Eq. \eqref{eq: prior tv}, respectively. The schematic diagram of our proposed F2PAD is illustrated in Fig. \ref{fig: schematic diagram}.

\subsection{Optimization Algorithm}
\label{sec: algorithm}

\begin{figure}[!t]
\centering 
\includegraphics[width=3.0 in]{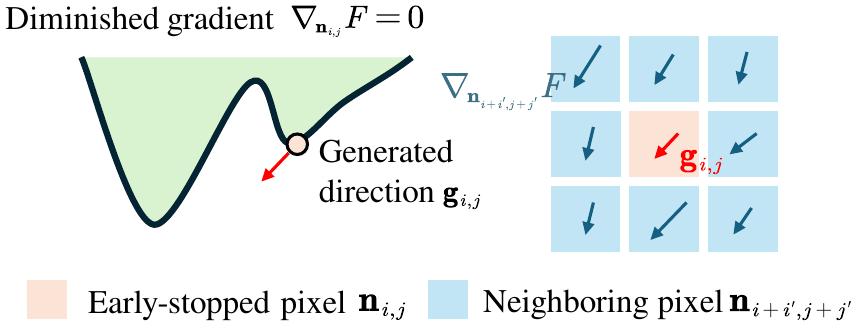}
\caption{Motivation of our proposed local gradient-sharing mechanism.} 
\label{fig: gradient}
\end{figure}

According to our discussions in Section \ref{sec: loss} and Eqs. \eqref{eq: sparse}, \eqref{eq: prior gmm}, and \eqref{eq: prior tv}, it is possible to use the gradient-based algorithm to solve the problem \eqref{eq: final model}. However, due to the complicated structure of \eqref{eq: final model} built on deep neural networks, the solutions of some pixels obtained by popular algorithms may be poor. The gradients will diminish at these pixels and thus early stop. To address this challenge, we propose a novel local gradient-sharing mechanism, as illustrated in Fig. \ref{fig: gradient}.

The key idea is that for a pixel, it is possible to use its spatially neighboring pixels' gradients to construct a feasible descending direction. In this way, the early-stopped pixels can still be pushed to escape the local minima. To achieve this, we can update the pixel $\mathbf{n}_{i,j}$ along the direction $\mathbf{g}_{i,j} \in \mathbb{R}^3$:
\begin{equation}
    \mathbf{g}_{i,j} = \sum_{i^{\prime},j^{\prime}} w_{i^{\prime},j^{\prime}} \cdot \nabla_{\mathbf{n}_{i+i^{\prime},j+j^{\prime}}}F(\mathbf{n}),
\label{eq: descending direction}
\end{equation}
where $i^{\prime},j^\prime \in [-k_s,k_s]$ and $k_s$ indicates the neighborhood size. $\nabla_{\mathbf{n}_{i+i^{\prime},j+j^{\prime}}}F(\mathbf{n})$ is the gradient of the objective function $F(\mathbf{n})$ w.r.t. the neighboring pixel $\mathbf{n}_{i+i^{\prime},j+j^{\prime}}$ in Eq. \eqref{eq: final model}. The weight $ w_{i^{\prime},j^{\prime}} \in \mathbb{R}$ is defined as:
\begin{equation}
    w_{i^{\prime},j^{\prime}} \propto \exp(-\dfrac{{i^{\prime}}^2+{j^{\prime}}^2}{\sigma_0})\exp(-\dfrac{\|\mathbf{x}_{i+i^{\prime},j+j^{\prime}}-\mathbf{x}_{i,j}\|_2^2}{\sigma_1}),
\label{eq: weights}
\end{equation}
where the two terms measure the influence of neighboring pixels in terms of their spatial locations and colors with hyperparameters $\sigma_0$ and $\sigma_1$ respectively. Eq. \eqref{eq: weights} indicates that we assign larger weights to the more similar pixels. Finally, we normalize $\{w_{i^\prime,j^\prime}\}$ such that $\sum_{i^{\prime},j^{\prime}}w_{i^{\prime},j^{\prime}}=1$.

In addition, our empirical results show that as $\mathbf{a}_{i,j} \rightarrow 0$, the corresponding gradient is relatively small, requiring a large step size to guarantee significant updating. Therefore, we propose an adaptive step size $\boldsymbol{\gamma}_{i,j} \in \mathbb{R}^3$ for the pixel $\mathbf{n}_{i,j}$ as:
\begin{equation}
    \boldsymbol{\gamma}_{i,j} = \gamma_0 / |\mathbf{a}_{i,j}| = \gamma_0 / |\mathbf{x}_{i,j}-\mathbf{n}_{i,j}|,
\label{eq: step size}
\end{equation}
where $\gamma_0$ is a constant and $|\cdot|$ takes the absolute values. Notably, we calculate $\boldsymbol{\gamma}_{i,j}$ in each iteration during optimization.

Finally, based on the descending direction $\mathbf{g}_{i,j}$ in Eq. \eqref{eq: descending direction} and step size $\boldsymbol{\gamma}_{i,j}$ in Eq. \eqref{eq: step size}, we update $\mathbf{n}_{i,j}$ following the rule in the adaptive Nesterov momentum algorithm (Adan) \cite{Xie2022AdanAN}. We will demonstrate the performance of our proposed algorithm compared to the standard Adan in Section \ref{sec: ablation algorithm}.

\subsection{Implementation}
\label{sec: implementation}

\begin{algorithm}[!t]
\caption{\small. The whole procedure of our F2PAD.}
\label{alg: whole}
\textbf{Input:} $\alpha_1$, $\alpha_2$, $\beta$, $\gamma_0$, $\sigma_0$, $\sigma_1$, $l_n(\mathbf{n})$, $\epsilon$, $\{\boldsymbol{\mu}_q,\boldsymbol{\mathbf{\Lambda}}_q\}$. \\
\textbf{Initialize}: $\mathbf{m}^{(0)}$. \\
\textbf{Output}: $\mathbf{m}^{(*)}$, and optional $\mathbf{n}^{(*)}$.

\begin{algorithmic}[1]
    \STATE Obtain the initial $\mathbf{n}^{(0)}$ by $\mathbf{m}^{(0)}$, as discussed in Section \ref{sec: initialization}.
    \STATE Obtain a sub-region of $\mathbf{n}$ to be optimized and construct the associated $l_n(\mathbf{n})$, according to Section \ref{sec: acceleration}.
    \STATE Calculate the weights $\{w_{i^{\prime},j^{\prime}}\}$ by Eq. \eqref{eq: weights}.
    \REPEAT
        \STATE Compute the descending directions $\{\mathbf{g}_{i,j}\}$ by Eq. \eqref{eq: descending direction}.
        \STATE Adjust the step sizes $\{\boldsymbol{\mathbf{\gamma}}_{i,j}\}$ by Eq. \eqref{eq: step size}.
        \STATE Update $\mathbf{n}$ by following Adan with $\{\mathbf{g}_{i,j}\}$ and $\{\boldsymbol{\mathbf{\gamma}}_{i,j}\}$.
    \UNTIL {Meet stopping criteria.}
\STATE Obtain the final $\mathbf{m}^{(*)}$ by Eqs. \eqref{eq: decomposition} and \eqref{eq: m}.
\STATE Re-estimate the final $\mathbf{n}^{(*)}$ according to $\mathbf{m}^{(*)}$, as discussed in Section \ref{sec: re-estimation}.
\end{algorithmic}
\end{algorithm}

This subsection presents the initialization and acceleration of our proposed framework and further summarizes the entire procedure.

\subsubsection{Initialization}
\label{sec: initialization}
To solve the problem \eqref{eq: final model}, it is required to obtain an initial solution $\mathbf{n}^{(0)}$.
Fortunately, we can utilize the anomaly mask $\mathbf{m}^{(0)}$ obtained by the original procedure of backbone methods. Concretely, the task of obtaining $\mathbf{n}^{(0)}$ can be transformed into an image inpainting task, i.e., infer the missing pixels under the masked region of $\mathbf{m}^{(0)}$. This paper utilizes a pretrained masked autoencoder (MAE) \cite {he2022masked} for inpainting. The MAE captures both local and global contextual relationships, capable of dealing with various sizes of missing regions. 
Notably, the MAE is expected to provide only a rough pixel-level reconstruction, sufficient for initialization. We will show the improvement of our final result by solving \eqref{eq: final model} compared to the initialization in Section \ref{sec: ablation initialization}.

\subsubsection{Acceleration}
\label{sec: acceleration}
To speed up our algorithm, we only optimize the region of $\mathbf{n}$ indicated as anomalous by  $\mathbf{m}^{(0)}$. As  $\mathbf{m}^{(0)}$ may not completely cover the ground truth (GT) anomaly, we enlarge $\mathbf{m}^{(0)}$ by simple morphological dilation. In addition, to calculate the loss $l_n(\mathbf{n})$, we only need to consider the subset of anomaly scores $\{s_{k,i,j}\}$ affected by the region of $\mathbf{n}$ being optimized.

\subsubsection{Re-estimation}
\label{sec: re-estimation}

Notably, we have obtained the final anomaly mask $\mathbf{m}^{(*)}$ by solving \eqref{eq: final model} and using Eq. \eqref{eq: m}, thereby completing the anomaly segmentation task. On the other hand, due to the relaxation of $l_0$ norm by the LOG penalty $l_a(\mathbf{a})$ in Eq. \eqref{eq: sparse}, the recovered $\mathbf{n}^{(*)}$ is always imperfect, which may be undesired in other applications. Hence, the re-estimation of $\mathbf{n}^{(*)}$ is required. Concretely, we can solve \eqref{eq: final model} again without the term $l_a(\mathbf{a})$ by setting $\beta=0$ and only optimizes the region within $\mathbf{m}^{(*)}$.

Finally, a small ($3\times3$) morphological open operator is applied to clean dot-like noises in the final mask, as suggested in \cite{salehi2021multiresolution}. 
We summarize our entire procedure in Algorithm \ref{alg: whole}, which terminates if it converges or reaches the maximum number of iterations.

\section{Case Study}
\label{sec: experiment}

This section provides case studies to demonstrate the effectiveness of our F2PAD. Specifically, the experiment descriptions and numerical results are summarized in Sections \ref{sec: description} and \ref{sec: main results} respectively. In addition, ablation studies are provided in Section \ref{sec: ablation} to investigate the influence of the important modules and steps in our method.

\subsection{Description}
\label{sec: description}

\begin{figure}[!t] 
\centering 
\includegraphics[width=0.49\textwidth]{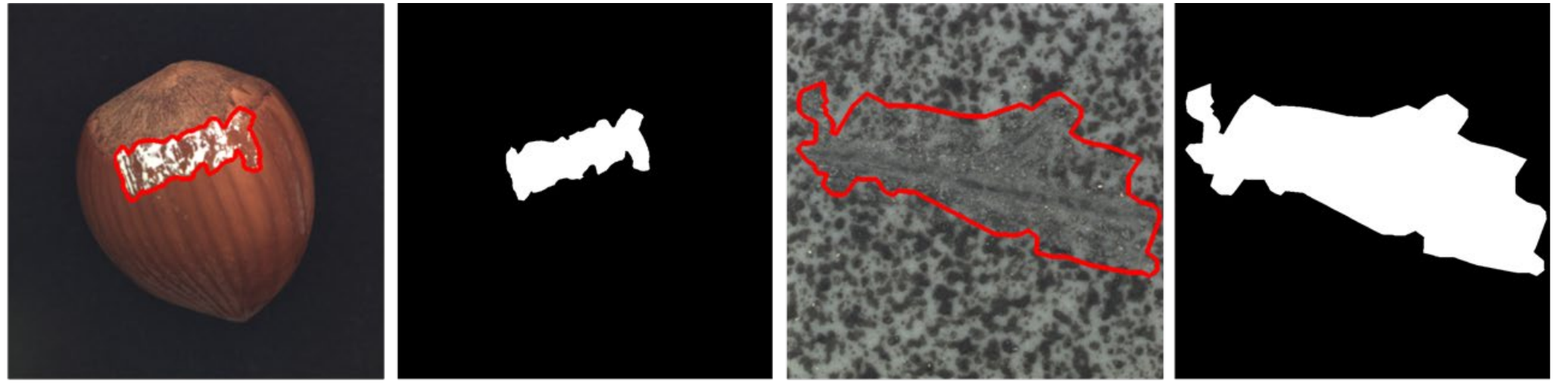} 
\caption{Examples of the inaccurate anomaly annotation in the popular MVTec AD dataset\cite{bergmann2019mvtec}.} 
\label{fig: annotation issue}
\end{figure}

\begin{figure}[!t] 
\centering 
\includegraphics[width=0.45\textwidth]{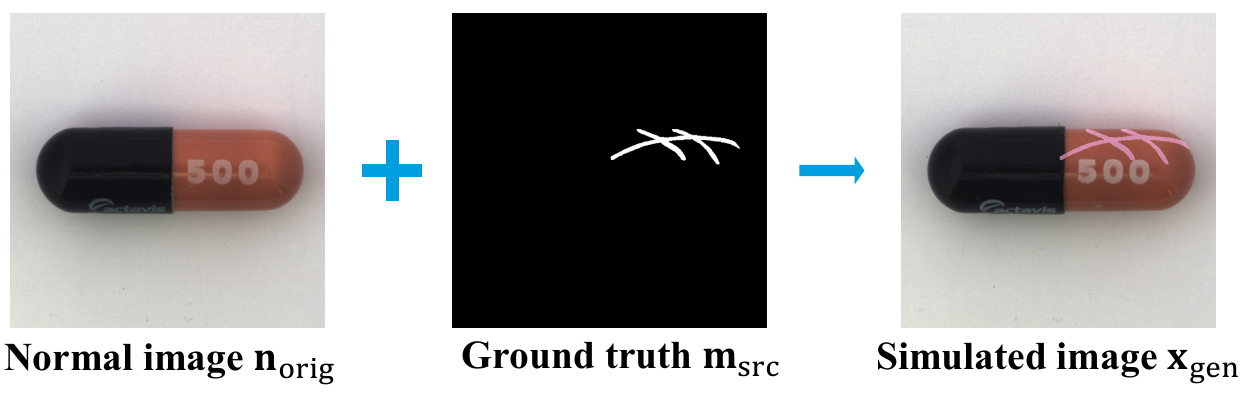} 
\caption{Illustration of anomaly generation by cut-paste.} 
\label{fig: flow dataset}
\end{figure}

\subsubsection{Dataset}
\label{sec: dataset} 
Many popular datasets suffer from inaccurate pixel-level annotations of GTs \cite{mou2023rgi}. For example, Fig.  \ref{fig: annotation issue} illustrates two examples in the MVTec AD dataset \cite{bergmann2019mvtec}, where the GTs cover more normal regions. To avoid it, this paper creates a new dataset based on the MVTec AD dataset to ensure precise annotations. Following the cut-paste strategy \cite{bauer2022self,bae2023pni}, which is widely adopted for anomaly generation, we first randomly pick a normal image $\mathbf{n}_{\mathrm{orig}}$ from the original MVTec AD dataset. Then, a defective image $\mathbf{x}_{\mathrm{gen}}$ is generated as:
\begin{equation}
    \mathbf{x}_{\mathrm{gen}} = (\mathbf{1}-\mathbf{m}_{\mathrm{src}}) \odot \mathbf{n}_{\mathrm{orig}} + \mathbf{m}_{\mathrm{src}} \odot \mathbf{a}_{\mathrm{src}},
\end{equation}
where $\odot$ indicates the element-wise multiplication, $\mathbf{a}_{\mathrm{src}}$ and $\mathbf{m}_{\mathrm{src}}$ are anomaly source and mask source respectively. Fig. \ref{fig: flow dataset} illustrates the above procedure.
For $\mathbf{a}_{\mathrm{src}}$, we simply select a random color \footnote{For the capsule category, we also introduce the structural anomaly. Specifically, we randomly select a region in the left(right) in another normal image and then paste it onto the right(left) part, as the capsule has a distinct left-right structure.}. $\mathbf{m}_{\mathrm{src}}$ is first selected from the MVTec AD dataset and then randomly resized. Notably, for the pixel $(i,j)$ with $\mathbf{m}_{\mathrm{src},i,j}=1$, it is required that $\mathbf{x}_{\mathrm{gen},i,j}$ and $\mathbf{n}_{\mathrm{orig},i,j}$ are also far from each other. If not satisfied, we change $\mathbf{x}_{\mathrm{gen},i,j}$ by multiplying a desired factor. 

Finally, three categories of textures (Wood, Leather, Tile) and objects (Hazelnut, Metal nut, Capsule) are created. The number of training samples ranges from 165 to 337, and 54 testing samples are generated for each category.

\subsubsection{Evaluation Metrics}
\label{sec: metric}

\begin{figure*}[htp]
\centering 
\includegraphics[width= 1\textwidth]{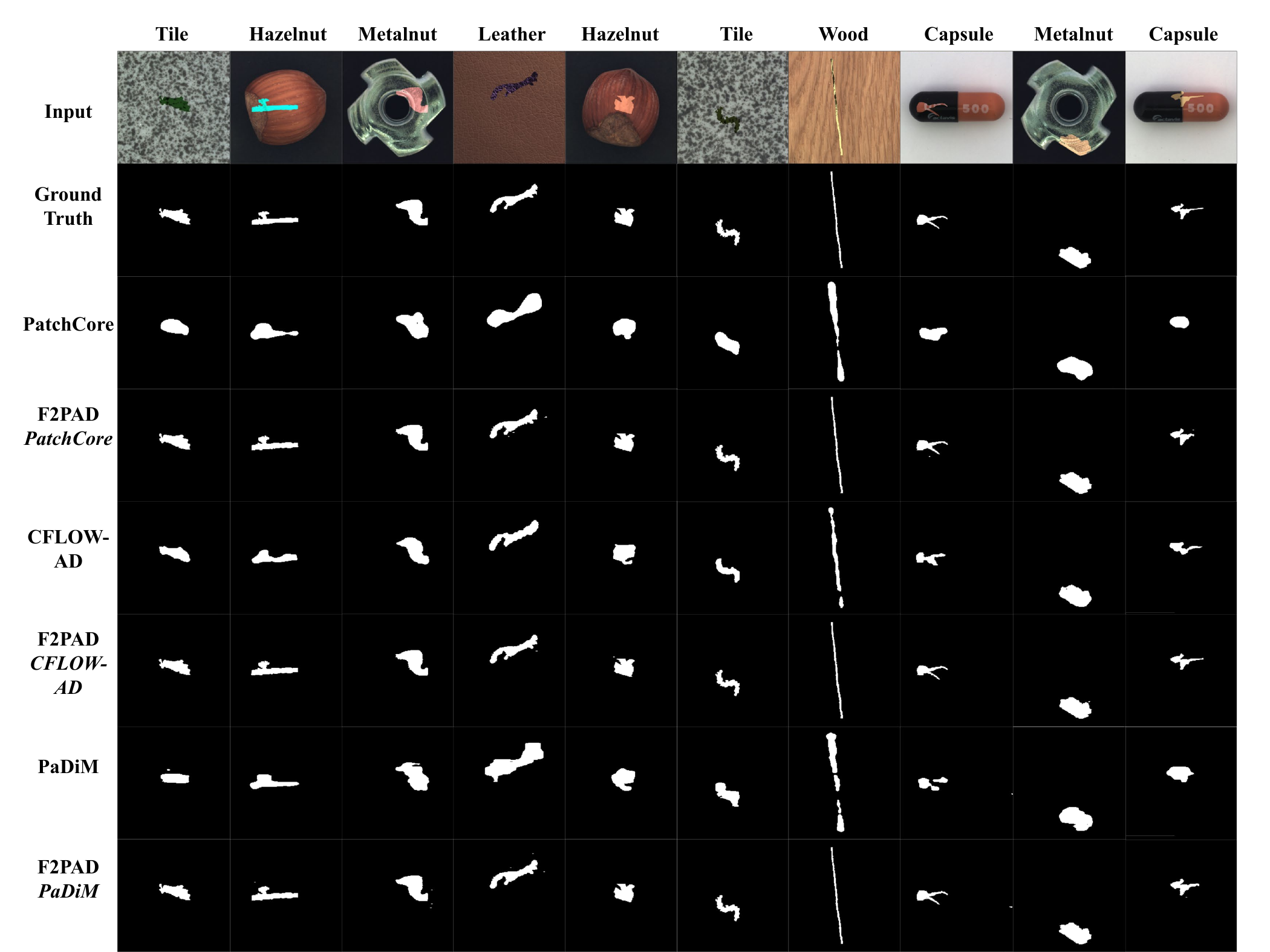}
\caption{Visualization of comparative results. Each row from top to bottom represents the input image, GT, the detection results of PatchCore, F2PAD based on PatchCore, CFLOW-AD, F2PAD based on CFLOW-AD, PaDiM, F2PAD based on PaDiM, respectively.} 
\label{fig: visualization result}
\end{figure*}

To better evaluate the accuracy of the anomaly boundaries detected by these methods, we choose the Intersection over Union (IOU) and DICE coefficient to evaluate our F2PAD, which are widely adopted in image segmentation tasks:
\begin{equation}
 \mathrm{DICE} =2\frac{|\mathbf{m} \cap \mathbf{m}_{\mathrm{gt}}|}{|\mathbf{m}|+|\mathbf{m}_{\mathrm{gt}}|}, \  \mathrm{IOU}=\frac{ |\mathbf{m} \cap \mathbf{m}_{\mathrm{gt}}|}{ |\mathbf{m} \cup \mathbf{m}_{\mathrm{gt}}|},
\label{eq:ioudice}
\end{equation}
where $\mathbf{m}_{\mathrm{gt}}$ is the GT mask. Furthermore, we multiply IOU and DICE by 100 for better illustration. Notably, higher IOU and DICE indicate better results of pixel-level anomaly segmentation.

\begin{table*}[htp]
    \centering
    \caption{Comparative results of our F2PAD and original backbone methods in the normal setting ($\Delta$: The difference in a metric between F2PAD and the original backbone).}
    \begin{tabular}{>{\centering\arraybackslash}p{1.2cm} 
    >{\centering\arraybackslash}p{0.4cm} >{\centering\arraybackslash}p{0.4cm} 
    >{\centering\arraybackslash}p{0.4cm}>{\centering\arraybackslash}p{0.4cm}
    >{\centering\arraybackslash}p{0.5cm}>{\centering\arraybackslash}p{0.5cm}
    >{\centering\arraybackslash}p{0.4cm} >{\centering\arraybackslash}p{0.4cm}
    >{\centering\arraybackslash}p{0.4cm}>{\centering\arraybackslash}p{0.4cm}
    >{\centering\arraybackslash}p{0.5cm}>{\centering\arraybackslash}p{0.5cm}
    >{\centering\arraybackslash}p{0.4cm} >{\centering\arraybackslash}p{0.4cm}
    >{\centering\arraybackslash}p{0.4cm} >{\centering\arraybackslash}p{0.4cm}
    >{\centering\arraybackslash}p{0.5cm}>{\centering\arraybackslash}p{0.5cm}}
    
     \hline \hline 
    \multirow{2}{*}{\textbf{Category}} &\multicolumn{2}{c}{\textbf{PatchCore}} &\multicolumn{4}{c}{\textbf{F2PAD \textit{PatchCore}}}&\multicolumn{2}{c}{\textbf{CFLOW-AD}} &\multicolumn{4}{c}{\textbf{F2PAD \textit{CFLOW-AD}}}&\multicolumn{2}{c}{\textbf{PaDiM}}&\multicolumn{4}{c}{\textbf{F2PAD \textit{PaDiM}}}\\ \cline{2-19}
     &IOU &DICE &IOU &$\Delta$ &DICE &$\Delta$ &IOU &DICE &IOU &$\Delta$ &DICE &$\Delta$ &IOU &DICE &IOU &$\Delta$ &DICE &$\Delta$ \\ \hline
     
    \textbf{Hazelnut}&	67.5&	80.1&	\textbf{86.9}&	19.5$\uparrow$&	\textbf{92.5}&	12.4$\uparrow$&	76.0&	86.1&	\textbf{86.3}&	10.3$\uparrow$&	\textbf{91.8}&	5.7$\uparrow$&	65.3&	78.5&	\textbf{88.3}&	23.0$\uparrow$&	\textbf{93.6}&	15.1$\uparrow$ \\ \hline
\textbf{Leather} &	70.4&	82.1&	\textbf{91.9}&	21.4$\uparrow$&	\textbf{95.7}&	13.6$\uparrow$&	77.9&	87.4&	\textbf{91.9}&	14.0$\uparrow$&	\textbf{95.4}&	8.0$\uparrow$&	61.6&	75.6&	\textbf{93.2}&	31.6$\uparrow$&	\textbf{96.4}&	20.8$\uparrow$ \\ \hline
\textbf{Metal nut} &71.5& 83.0&	\textbf{89.4}&	17.9$\uparrow$&	\textbf{93.7}&	10.7$\uparrow$&	78.4&	87.4&	\textbf{89.3}&	10.9$\uparrow$&	\textbf{93.8}&	6.4$\uparrow$&	68.9&	81.3&	\textbf{91.1}&	22.2$\uparrow$&	\textbf{95.0}&	13.7$\uparrow$ \\ \hline
\textbf{Capsule}&	60.6&	74.6&	\textbf{86.1}&	25.5$\uparrow$&	\textbf{92.4}&	17.8$\uparrow$&	68.4&	80.8&	\textbf{83.4}&	15.0$\uparrow$&	\textbf{90.3}&	9.5$\uparrow$&	55.3&	70.4&	\textbf{85.5}&	30.2$\uparrow$&	\textbf{92.0}&	21.6$\uparrow$ \\ \hline
\textbf{Tile}&	69.2&	81.4&	\textbf{95.0}&	25.8$\uparrow$&	\textbf{97.4}&	15.9$\uparrow$&	87.1&	93.0&	\textbf{95.4}&	8.3$\uparrow$&	\textbf{97.1}&	4.1$\uparrow$&	64.5&	78.0&	\textbf{96.7}&	32.2$\uparrow$&	\textbf{98.3}&	20.3$\uparrow$ \\ \hline
\textbf{Wood}&	59.9&	73.5&	\textbf{86.6}&	26.7$\uparrow$&	\textbf{91.8}&	18.3$\uparrow$&	70.0&	81.8&	\textbf{90.5}&	20.5$\uparrow$&	\textbf{94.3}&	12.4$\uparrow$&	51.2&	66.6&	\textbf{87.5}&	36.2$\uparrow$&	\textbf{92.2}&	25.6$\uparrow$ \\ \hline
\textbf{Mean}&	66.5&	79.1&	\textbf{89.3}&	22.8$\uparrow$&	\textbf{93.9}&	14.8$\uparrow$&	76.3&	86.1&	\textbf{89.5}&	13.2$\uparrow$&	\textbf{93.8}&	7.7$\uparrow$&	61.1&	75.1&	\textbf{90.4}&	29.3$\uparrow$&	\textbf{94.6}&	19.5$\uparrow$ \\
     \hline \hline
    \end{tabular}
    \label{table: normal}
\end{table*}

\begin{table*}[!t]
 \centering
 \caption{Comparative results of our F2PAD and original backbone methods in the few-shot setting with only 16 images for training  ($\Delta$: The difference in a metric between F2PAD and the original backbone).}
 \begin{tabular}{>{\centering\arraybackslash}p{1.2cm} 
 >{\centering\arraybackslash}p{0.4cm} >{\centering\arraybackslash}p{0.4cm} 
    >{\centering\arraybackslash}p{0.4cm}>{\centering\arraybackslash}p{0.4cm}
    >{\centering\arraybackslash}p{0.5cm}>{\centering\arraybackslash}p{0.5cm}
    >{\centering\arraybackslash}p{0.4cm} >{\centering\arraybackslash}p{0.4cm}
    >{\centering\arraybackslash}p{0.4cm}>{\centering\arraybackslash}p{0.4cm}
    >{\centering\arraybackslash}p{0.5cm}>{\centering\arraybackslash}p{0.5cm}
    >{\centering\arraybackslash}p{0.4cm} >{\centering\arraybackslash}p{0.4cm}
    >{\centering\arraybackslash}p{0.4cm} >{\centering\arraybackslash}p{0.4cm}
    >{\centering\arraybackslash}p{0.5cm}>{\centering\arraybackslash}p{0.5cm}}
 
 \hline \hline 
 \multirow{2}{*}{\textbf{Category}} &\multicolumn{2}{c}{\textbf{PatchCore}} &\multicolumn{4}{c}{\textbf{F2PAD \textit{PatchCore}}}&\multicolumn{2}{c}{\textbf{CFLOW-AD}} &\multicolumn{4}{c}{\textbf{F2PAD \textit{CFLOW-AD}}}&\multicolumn{2}{c}{\textbf{PaDiM}}&\multicolumn{4}{c}{\textbf{F2PAD \textit{PaDiM}}}\\ \cline{2-19}
 &IOU &DICE &IOU &$\Delta$ &DICE &$\Delta$ &IOU &DICE &IOU &$\Delta$ &DICE &$\Delta$ &IOU &DICE &IOU &$\Delta$ &DICE &$\Delta$ \\ \hline
 
 \textbf{Hazelnut}&	66.9&	79.6&	\textbf{85.8}&	18.9$\uparrow$&	\textbf{92.2}&	12.6$\uparrow$&	67.3&	79.9&	\textbf{84.5}&	17.2$\uparrow$&	\textbf{90.7}&	10.8$\uparrow$&	\textbf{63.8}&	77.5&	\textbf{88.2}&	24.5$\uparrow$&	\textbf{92.7}&	15.3$\uparrow$ \\ \hline
\textbf{Leather}&	70.3&	82.1&	\textbf{90.9}&	20.6$\uparrow$&	\textbf{94.7}&	12.7$\uparrow$&	70.6&	82.5&	\textbf{91.2}&	20.6$\uparrow$&	\textbf{94.8}&	12.3$\uparrow$&	66.9&	79.7&	\textbf{89.5}&	22.7$\uparrow$&	\textbf{92.9}&	13.2$\uparrow$\\ \hline
\textbf{Metal nut}&	70.9&	82.6&	\textbf{91.1}&	20.2$\uparrow$&	\textbf{95.3}&	12.7$\uparrow$&	63.4&	76.6&	\textbf{87.5}&	24.1$\uparrow$&	\textbf{92.5}&	15.9$\uparrow$&	67.9&	80.5&	\textbf{88.8}&	20.8$\uparrow$&	\textbf{92.6}&	12.1$\uparrow$\\ \hline
\textbf{Capsule}&	60.1&	74.2&	\textbf{83.6}&	23.5$\uparrow$&	\textbf{90.3}&	16.2$\uparrow$&	44.9&	60.8&	\textbf{69.0}&	24.1$\uparrow$&	\textbf{78.2}&	17.5$\uparrow$&	57.1&	71.9&	\textbf{86.1}&	29.0$\uparrow$&	\textbf{92.4}&	20.5$\uparrow$ \\ \hline
\textbf{Tile}&	68.7&	81.1&	\textbf{94.8}&	26.1$\uparrow$&	\textbf{97.3}&	16.2$\uparrow$&	73.6&	84.5&	\textbf{93.4}&	19.7$\uparrow$&	\textbf{96.3}&	11.8$\uparrow$&	66.5&	79.5&	\textbf{89.1}&	22.6$\uparrow$&	\textbf{93.7}&	14.3$\uparrow$ \\ \hline
\textbf{Wood}&	59.7&	73.2&	\textbf{87.8}&	28.1$\uparrow$&	\textbf{92.3}&	19.1$\uparrow$&	63.2&	77.0&	\textbf{89.2}&	25.9$\uparrow$&	\textbf{93.4}&	16.4$\uparrow$&	57.4&	71.6&	\textbf{86.0}&	28.6$\uparrow$&	\textbf{91.3}&	19.7$\uparrow$ \\ \hline
\textbf{Mean}&	66.1&	78.8&	\textbf{89.0}&	22.9$\uparrow$&	\textbf{93.7}&	14.9$\uparrow$&	63.8&	76.9&	\textbf{85.8}&	22.0$\uparrow$&	\textbf{91.0}&	14.1$\uparrow$&	63.3&	76.8&	\textbf{88.0}&	24.7$\uparrow$&	\textbf{92.6}&	15.9$\uparrow$ \\ 
 \hline \hline
 \end{tabular}
 \label{table: few-shot}
\end{table*}

\subsubsection{Implementation Details}
\label{sec: implement details}

Each image is resized to $224\times224$ before model training and testing. Following the official implementations \footnote{{https://github.com/xiahaifeng1995/PaDiM-Anomaly-Detection-Localization-master; https://github.com/gudovskiy/cflow-ad}}, 
Resnet18 is chosen as the feature extractor for PaDiM and PatchCore, while EfficientNet-B6 is for CFLOW-AD. For PatchCore, $|\mathcal{M}_{i,j}| = 50$ for all $i,j$. For CFLOW-AD, we only consider the first two feature maps in the few-shot setting to avoid overfitting, see Section \ref{sec: few-shot}. For all backbone methods, we choose the optimal thresholds of anomaly scores that maximize F1 scores to obtain their binary anomaly masks, following \cite{defard2021padim}.
We fix $\epsilon=0.0001$ in $l_a({\mathbf{a}})$ and $\alpha^{2}=0.0001$ for $l_p^2({\mathbf{n}})$. 

Moreover, our empirical results show that a small $\beta$ in \eqref{eq: final model} is required for large anomalies. Hence, instead of adopting a constant $\beta$ for different sizes of anomalies, we set $\beta = \beta_0/\sum_{i,j}\mathbf{m}^{(0)}_{i,j}$, where the denominator indicates the estimated anomaly size by backbone methods. We fine-tune the hyperparameters $\alpha^1$ and $\beta_0$, along with the number of components in MOG, by grid-search for each category.

For our proposed algorithm in Section \ref{sec: algorithm}, $\gamma_0$, $k_s$, $\sigma_0$, and $\sigma_1$ are set as 1.0, 5, 1.1, and 3.0, respectively. Additionally, we apply the gradient clipping to $\textbf{g}_{i,j}$ by 0.03 to avoid too large values.
We use the recommended parameters of Adan \cite{Xie2022AdanAN}.  The optimization terminates when the loss decreases between two subsequent iterations less than 0.1 for PaDiM and 0.05 for other backbones, or running exceeds 1200 iterations. Additionally, we also penalize the pixel values to be valid during optimization. 

Finally, for our ablation study in Section \ref{sec: ablation initialization}, we obtain the binary mask of initialization by thresholding the pixel difference in the $l_2$ norm between the input and recovered images by inpainting at 0.2 \footnote{The pixel values are normalized using the mean and standard deviations of ImageNet images, adhering to the conventions of the computer vision community.}.

\subsection{Results}
\label{sec: main results}

\begin{figure}[!t]
\centering 
\includegraphics[width=3.4in]{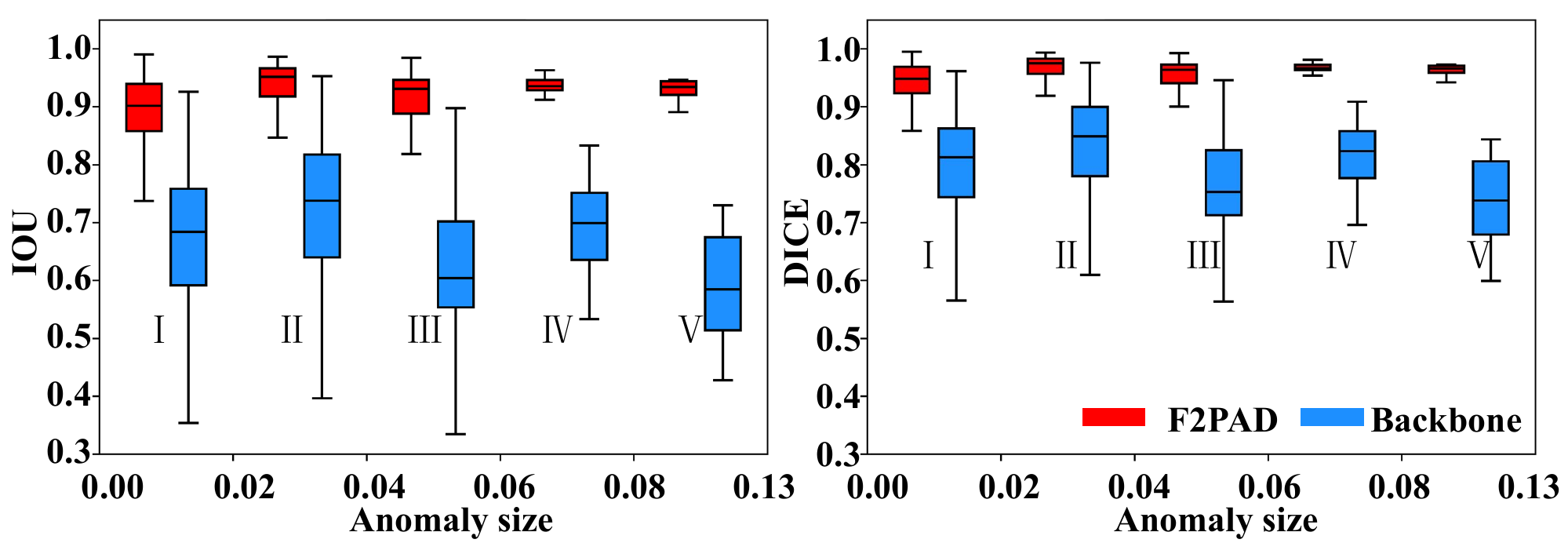}
\caption{Results of F2PAD (red) and all backbone methods (blue) across various anomaly sizes over the whole dataset. Group \uppercase\expandafter{\romannumeral1} includes samples with anomaly sizes ranging from 0 to 0.02, while Group \uppercase\expandafter{\romannumeral5} represents those with anomaly sizes above 0.08.} 
\label{fig: anomaly size}
\end{figure}

\subsubsection{Normal Setting}
\label{sec: normal}

Table \ref{table: normal} presents the results of our F2PAD when applied to PatchCore, PaDiM, and CFLOW-AD, demonstrating a significant improvement of IOU and DICE in all 6 categories. The columns of $\Delta$ in Table \ref{table: normal} show the differences in IOU and DICE between F2PAD and the original backbones. Moreover, we visualize the results of 10 representative samples in Fig. \ref{fig: visualization result}. Though the feature-based methods can localize the anomalies as illustrated in Fig. \ref{fig: visualization result}, the detected anomalies exhibit inaccurate boundaries with relatively low IOU and DICE. For example,  Table \ref{table: normal} shows that the original PatchCore method only achieves an average IOU of 66.5 points and DICE of 79.1 points respectively. In contrast, our F2PAD can obtain more accurate anomaly segmentation results as demonstrated by Fig. \ref{fig: visualization result} and Table \ref{table: normal}, improving the IOU and DICE of PatchCore to 89.3 points and 93.9 points respectively. Similarly, our F2PAD results in improvements of 13.2 and 7.7 points in IOU and DICE  for CFLOW-AD, respectively, and 29.3 and 19.5 points for PaDiM, respectively.
The improvements benefit from our model \eqref{eq: final model}, which directly handles \textbf{Issues 1\&2} of feature-based methods, as discussed in Section \ref{sec:introcution}. 

In addition, Fig. \ref{fig: anomaly size} summarizes the results across all categories and backbones concerning anomaly size.
It categorizes samples into five groups (\uppercase\expandafter{\romannumeral1} to \uppercase\expandafter{\romannumeral5}), based on anomaly sizes within the intervals from [0, 0.02] to [0.08, 0.13], respectively. Here, anomaly size refers to the ratio of the anomaly area to the total image area, with 0.13 being the largest anomaly size in our dataset.  Fig. \ref{fig: anomaly size} demonstrates the consistent improvements of our F2PAD over different sizes of anomalies.

\subsubsection{Few-Shot Setting}
\label{sec: few-shot}

One benefit of feature-based methods is their less demand for the sample size in training, particularly relevant in real manufacturing scenarios with small-batch production. Therefore, we provide the few-shot setting to validate the performance of our F2PAD under the above scenario. Specifically, we only use 16 samples for each category during training, as suggested in \cite{roth2022towards}, while the test dataset adopted in Section \ref{sec: normal} remains unchanged. The results are summarized in Table \ref{table: few-shot}. 
Due to the more challenging setting, the performance of the backbone methods decreases slightly, as shown in Table \ref{table: few-shot}. Nevertheless,  our F2PAD still achieves significant improvement in IOU and DICE for all backbones and categories. 

In conclusion, our F2PAD can successfully handle three backbones in various data categories and apply to normal and few-shot settings.

\subsection{Ablation Study}
\label{sec: ablation}
\begin{figure*}[!ht] 
\centering 
\includegraphics[width=1.0 \textwidth]{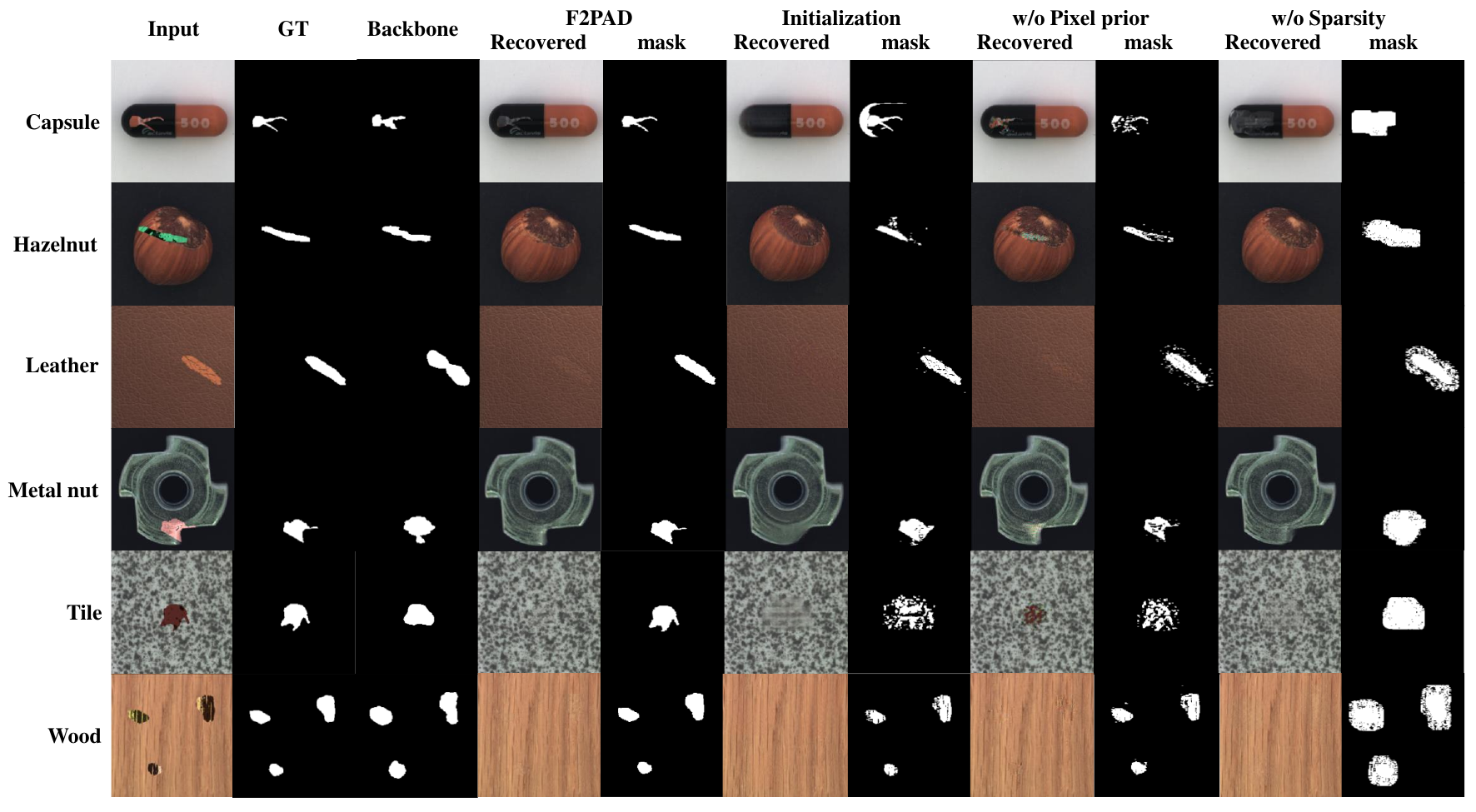} 
\caption{Visualization of the results of ablation study on initialization and regularization terms. Each column from left to right represents the input image, GT, the mask obtained by original backbone methods, the recovered non-defective images and masks obtained by F2PAD, initialization, F2PAD without the pixel prior $l_p(\mathbf{n})$, and F2PAD without the sparsity-inducing penalty $l_a(\mathbf{a})$, respectively. } 
\label{fig: ablation study}
\end{figure*}

\begin{table}[!t]
 \centering
 \caption{Ablation study of initialization and regularization over the whole dataset.}
 \begin{tabular}{
 >{\centering\arraybackslash}p{2.0cm} 
 >{\centering\arraybackslash}p{0.5cm} 
 >{\centering\arraybackslash}p{0.5cm} 
 >{\centering\arraybackslash}p{0.5cm}
 >{\centering\arraybackslash}p{0.5cm}
 >{\centering\arraybackslash}p{0.5cm}
 >{\centering\arraybackslash}p{0.5cm}
    }
 \hline \hline 
 \multirow{2}{*}{\textbf{Ablation}} 
 &\multicolumn{2}{>{\centering\arraybackslash}p{1.6cm}}{\textbf{F2PAD \textit{PatchCore}}}
 &\multicolumn{2}{>{\centering\arraybackslash}p{1.6cm}}{\textbf{F2PAD \textit{CFLOW-AD}}}
 &\multicolumn{2}{>{\centering\arraybackslash}p{1.6cm}}{\textbf{F2PAD \textit{PaDiM}}}\\ \cline{2-7}
 &IOU &DICE &IOU  &DICE  &IOU &DICE \\ \hline
 \textbf{Initialization} &71.9 &82.0 &71.9 &82.1 &71.9 &82.0\\ \hline
 \textbf{w/o Pixel prior} &72.6 &82.0& 71.4&	80.8&69.9&79.7 \\ \hline
\textbf{w/o Sparsity}& 32.6&48.6& 32.5&	48.5& 28.8 & 44.2 \\ \hline
\textbf{F2PAD} &\textbf{89.3} &\textbf{93.9} &\textbf{89.5} &\textbf{93.8} &\textbf{90.4} &\textbf{94.6}\\
\hline
 \hline 
 \end{tabular}
 \label{table: ablation}
\end{table}

\subsubsection{Comparison with Initialization}
\label{sec: ablation initialization}

We first demonstrate the effectiveness of our optimization model \eqref{eq: final model}, compared to the initial solution obtained by inpainting as discussed in Section \ref{sec: initialization}. The results of IOU and DICE are presented in Table \ref{table: ablation}, where the final solution of \eqref{eq: final model} improves IOU and DICE by above 17.4 points (PatchCore) and 11.7 points (CFLOW-AD) respectively. Furthermore, Fig. \ref{fig: ablation study} showcases visualized results for six samples across diverse categories. The recovered images of initialization lose certain textural details. For example, the random spots of Tile are heavily obscured, as shown in Fig. \ref{fig: ablation study}. Moreover, the initialization may also generate incompatible structures with target objects. For example, the bottom tooth of the Metal nut in Fig. \ref{fig: ablation study} is significantly warped. The reason for the above phenomena is that we adopt the large pretrained model for image inpainting to obtain the initial solution, it lacks sufficient domain knowledge to ensure precise reconstruction of unseen objects in manufacturing 
scenarios. Therefore, the inaccuracies at the pixel level are inevitable. However, the initial solution can initialize the further optimization of our F2PAD, leading to satisfactory pixel-level anomaly segmentation, as illustrated in Figs. \ref{fig: visualization result} and \ref{fig: ablation study}.

\subsubsection{Importance of Regularization}
\label{sec: ablation regularization}

\begin{figure}[!t] 
\centering 
\includegraphics[width=0.49 \textwidth]{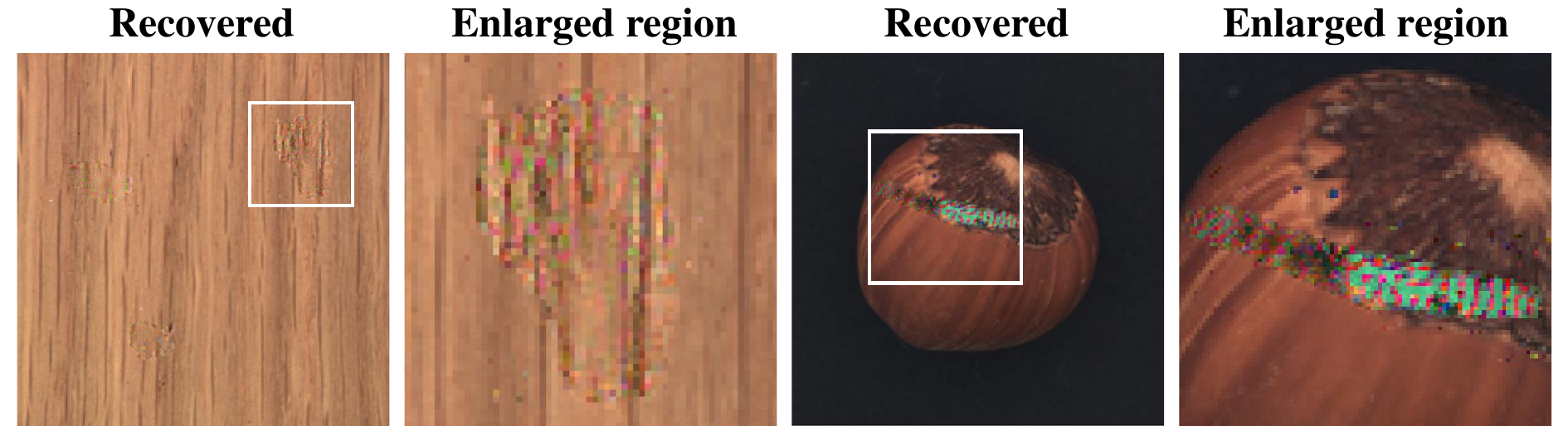} 
\caption{Illustration of the corrupted pixels in the recovered images (Wood and Hazelnut) of Fig. \ref{fig: ablation study} without the pixel prior term, where we enlarge the regions within the white rectangles in the right columns. }
\label{fig: pixel prior}
\end{figure}

This part discusses the importance of the pixel prior term $l_p(\mathbf{n})$ and the sparsity-inducing penalty $l_a(\mathbf{a})$ in our model \eqref{eq: model}. Concretely, we remove $l_p(\mathbf{n})$ and $l_n(\mathbf{n})$ respectively, denoted as w/o Pixel prior and w/o Sparsity in Table \ref{table: ablation} and Fig. \ref{fig: ablation study}. Table \ref{table: ablation} shows a decrease of IOU and DICE if removing the two regularization terms.

As discussed in Section \ref{sec: pixel prior}, the pixel prior $l_p(\mathbf{n})$ plays an important role in recovering the pixel values. As illustrated in Fig. \ref{fig: ablation study}, the anomaly masks generated without incorporating pixel priors are incomplete, with discernible missing areas within their interior regions. To understand it, Fig. \ref{fig: pixel prior} enlarges the anomalous regions of recovered images, showing heavily corrupted pixels similar to the color noise. Although these corrupted pixels may not distort the semantic information too much, the integrity of anomaly masks at the pixel level will deteriorate. Moreover, the uncertainties associated with the corrupted pixels also complicate the optimization process, making it more challenging and unstable. In contrast, by incorporating the pixel prior, our F2PAD
obtains more complete anomaly masks and realistic non-defective images, as shown in Fig. \ref{fig: ablation study}.

Finally, the sparsity-inducing penalty $l_a(\mathbf{a})$ enables accurate anomaly segmentation by restricting the locality of anomalies. As shown in Fig. \ref{fig: ablation study}, the results without sparsity cannot identify the anomaly boundaries, which cover more normal regions and thus result in the lowest IOU and DICE in Table \ref{table: ablation}.

\subsubsection{Performance of Optimization Algorithm}
\label{sec: ablation algorithm}

\begin{figure}[!t] 
\centering 
\includegraphics[width=0.49 \textwidth]{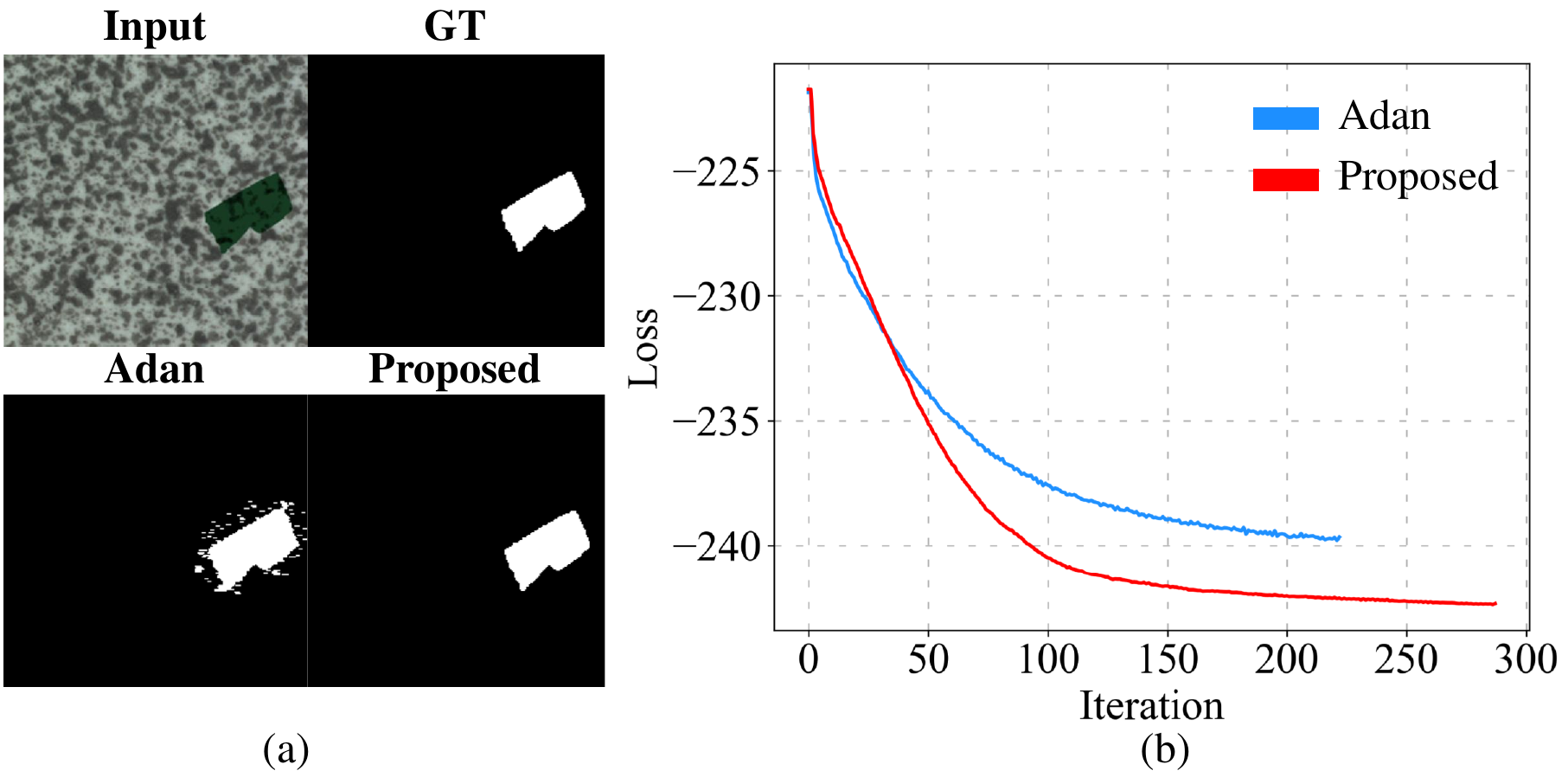} 
\caption{ Visualization of the results obtained by our proposed optimization algorithm and standard Adan \cite{Xie2022AdanAN} with an example from the Tile category when applied to PatchCore. (a) From top-left to bottom-right, the input image, GT, and masks are displayed, respectively. (b) Loss trajectories during the optimization process.} 
\label{fig: ablation optimization}
\end{figure}

We show the superiority of our proposed optimization algorithm in Section \ref{sec: algorithm} to the original Adan \cite{Xie2022AdanAN}. We provide an example from the Tile category in Fig. \ref{fig: ablation optimization} when applied to PatchCore. The bottom-left plot in Fig. \ref{fig: ablation optimization} (a) illustrates that some normal pixels are mislabeled in the result obtained by the original Adan. This observation indicates that these pixels are early-stopped during optimization, consistent with the higher loss value in Fig. \ref{fig: ablation optimization} (b). In contrast, thanks to the local gradient-sharing mechanism in Section \ref{sec: ablation algorithm}, our proposed algorithm converges into a lower loss value and thus provides a more accurate anomaly mask, as shown in Fig. \ref{fig: ablation optimization}.

Overall, both our model as defined in \eqref{eq: model}, the regularization terms, and the proposed optimization algorithm are critical for our method to enable precise pixel-level anomaly segmentation.

\subsection{Discussion}

\begin{table}[!t]
 \centering
 \caption{The average number of iterations and running time per sample during optimization in the normal setting.}
 \begin{tabular}{
 >{\centering\arraybackslash}p{2.0cm} 
 >{\centering\arraybackslash}p{1.5cm} 
 >{\centering\arraybackslash}p{1.5cm} 
 >{\centering\arraybackslash}p{1.5cm}
    }
 \hline \hline 
\multirow{2}{*}{\textbf{Item}}
 &\textbf{F2PAD \textit{PatchCore}}
 &\textbf{F2PAD \textit{CFLOW-AD}}
 &\textbf{F2PAD \textit{PaDiM}}\\ \cline{1-4}
\textbf{\# Iteration } &202.3 &204.0 &398.1 \\ \hline
\textbf{Time (Seconds)} & 17.8 & 8.7 & 40.8 \\
\hline
 \hline 
 \end{tabular}
 \label{table: running time}
\end{table}

\subsubsection{Time Analysis}
We summarize the running time of our F2PAD to demonstrate its practicability. Specifically, we implemented our F2PAD on Python 3.9 with CPU Intel i7 10700, GPU GTX 3080 10GB, and RAM 64GB. The average number of iterations until convergence and total time of solving the problem \eqref{eq: final model} in the normal setting in Section \ref{sec: normal} are reported in Table \ref{table: running time}. 
With a minimal average total time per sample of 8.7 seconds for F2PAD \textit{CFLOW-AD} and a maximum of 40.8 seconds for F2PAD \textit{PaDiM}, our F2PAD demonstrates good applicability to real-world quality control scenarios. Notably, the relatively low computational speed of F2PAD \textit{PaDiM} is related to the large number of $\{\mathbf{\mu}_{i,j}\}$ and $\{\mathbf{\Lambda}_{i,j}\}$ required for each location $i,j$ in Eq. \eqref{eq: PaDiM}. One may consider to approximate $\{\mathbf{\mu}_{i,j}\}$ and $\{\mathbf{\Lambda}_{i,j}\}$ with fewer bases, which remains our future study. 

\subsubsection{Failure Analysis}

\begin{figure}[!t] 
\centering 
\includegraphics[width=0.45 \textwidth]{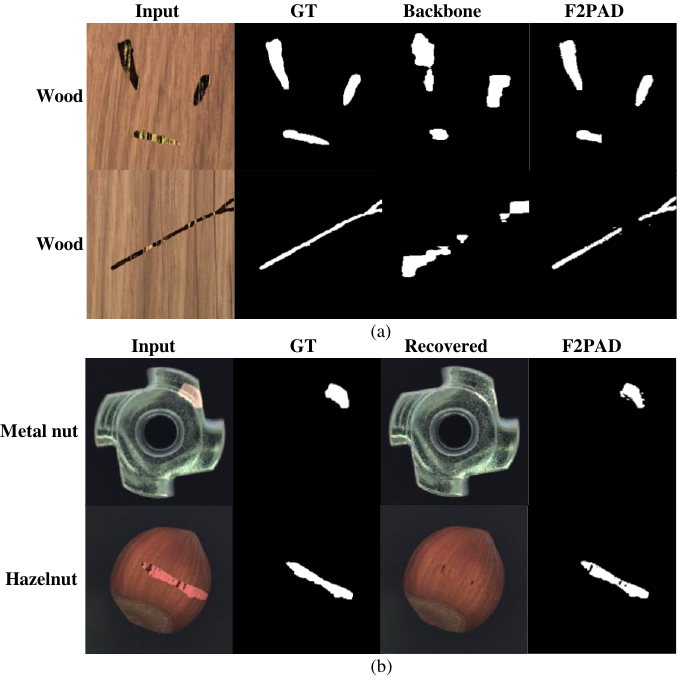} 
\caption{Visualization of failure scenarios. (a) Cases in which the initial masks, obtained through the backbone method, are significantly incomplete. (b) Cases in which specific regions of anomalies appear ambiguous, and the associated areas in the images recovered by our F2PAD exhibit properties akin to those of normal areas.}
\label{fig: failure}
\end{figure}

Fig. \ref{fig: failure} summarizes two types of failure cases in our experiment. The first case in Fig. \ref{fig: failure}(a) stems from incomplete masks generated by backbone methods. To initialize our algorithm (see Section \ref{sec: acceleration}), we concentrate on areas within dilated masks, which may occasionally fail to cover the GTs. The second case, illustrated in Fig. \ref{fig: failure}(b), occurs when our F2PAD fails to recover some anomalous pixels. These pixels are visually compatible with the recovered images, yielding appearances nearly indistinguishable from normal ones. Nevertheless, these failed cases are rare in our experiment.

\section{Conclusion}
\label{sec: conclusion}

Unsupervised image anomaly detection is critical for various manufacturing applications, where precise pixel-level anomaly segmentation is required for downstream tasks. However, existing feature-based methods show inaccurate anomaly boundaries due to the decreased resolution of the feature map and the mixture of adjacent normal and anomalous pixels. To address the above issues, we proposed a novel optimization framework, namely F2PAD, to enhance various feature-based methods, including the popular PatchCore, CFLOW-AD, and PaDiM. Our extensive case studies demonstrated that our F2PAD improved remarkably over three backbone methods in normal and few-shot settings. Moreover, the necessity of each component in F2PAD was demonstrated by our ablation studies. Our F2PAD is suitable for handling realistic and complex manufacturing scenarios, such as small-batch manufacturing. It has the potential to become a general framework to enhance feature-based anomaly detection methods.


%





\ifCLASSOPTIONcaptionsoff
  \newpage
\fi





\bibliographystyle{IEEEtran}
\bibliography{IEEEabrv,Bibliography}

\end{document}